\documentclass[%
superscriptaddress,
preprint,
showpacs,
amsmath,amssymb,
prc,
floatfix, ]%
{revtex4-1}
\usepackage{}
\usepackage{mathrsfs}

\usepackage{color}
\usepackage{CJK}

\usepackage{graphicx}% Include figure files
\usepackage{dcolumn}% Align table columns on decimal point
\usepackage{bm}% bold math
\usepackage[dvipdfm,bookmarks=true,colorlinks,%
            citecolor=blue,linkcolor=blue,hypertex, %
            breaklinks=true]{hyperref}

\begin{document}

%\begin{CJK*}{GBK}{}

%\preprint{NP-ITP/CAS-201002}

\title{Nuclear superfluidity for antimagnetic
rotation in $^{105}$Cd and $^{106}$Cd}

\author{Zhen-Hua Zhang}%
 \affiliation{State Key Laboratory of Nuclear Physics and Technology,
              School of Physics, Peking University, Beijing 100871, China}
\author{Peng-Wei Zhao }%
 \affiliation{State Key Laboratory of Nuclear Physics and Technology,
              School of Physics, Peking University, Beijing 100871, China}
\author{Jie Meng }%
% \email{mengj@pku.edu.cn}
 \affiliation{State Key Laboratory of Nuclear Physics and Technology,
              School of Physics, Peking University, Beijing 100871, China}
 \affiliation{School of Physics and Nuclear Energy Engineering,
              Beihang University, Beijing 100191, China}
 \affiliation{Department of Physics, University of Stellenbosch,
              Stellenbosch 7602, South Africa}
\author{Jin-Yan Zeng }%
 \affiliation{State Key Laboratory of Nuclear Physics and Technology,
              School of Physics, Peking University, Beijing 100871, China}
\author{En-Guang Zhao}%
 \affiliation{State Key Laboratory of Nuclear Physics and Technology,
              School of Physics, Peking University, Beijing 100871, China}
 \affiliation{State Key Laboratory of Theoretical Physics,
              Institute of Theoretical Physics, Chinese Academy of Sciences,
              Beijing 100190, China}
 \affiliation{Center of Theoretical Nuclear Physics, National Laboratory
              of Heavy Ion Accelerator, Lanzhou 730000, China}
\author{Shan-Gui Zhou}
 \affiliation{State Key Laboratory of Theoretical Physics,
              Institute of Theoretical Physics, Chinese Academy of Sciences,
              Beijing 100190, China}
 \affiliation{Center of Theoretical Nuclear Physics, National Laboratory
              of Heavy Ion Accelerator, Lanzhou 730000, China}

\date{\today}

\begin{abstract}
The effects of nuclear superfluidity on antimagnetic rotation bands
in $^{105}$Cd and $^{106}$Cd are investigated by the cranked shell
model with the pairing correlations and the blocking effects
treated by a particle-number conserving method.
The experimental moments of inertia and the reduced $B(E2)$
transition values are excellently reproduced.
The nuclear superfluidity is essential to reproduce the
experimental moments of inertia.
The two-shears-like mechanism for the antimagnetic rotation is
investigated by examining the shears angle, i.e., the closing of the
two proton hole angular momenta, and its sensitive dependence on the
nuclear superfluidity is revealed.
\end{abstract}

\pacs{21.60.-n; 21.10.Re; 23.20-g; 27.60.+j}%

%21.60.-n 	Nuclear structure models and methods
%21.10.Re 	Collective levels
%23.20-g 	Electromagnetic transitions
%27.60.+j   90¡Ü A ¡Ü 149
\maketitle

\section{\label{Sec:Introduction}Introduction}

Most rotational bands are usually observed in nuclei
with substantial quadrupole deformations.
In these bands, the states decay by strong electric quadrupole ($E2$)
transitions and the energy spectra show a pronounced rotational character.
Such bands are usually interpreted as the result of a coherent
collective rotation of many nucleons around an axis perpendicular
to the symmetry axis.

In the 1990's, however, a new type of rotational band with
strongly enhanced magnetic dipole ($M1$) transitions and
very weak $E2$ transitions has been discovered experimentally
in nearly spherical light Pb isotopes~\cite{Hubel2005_PPNP54-1}.
This new type of rotational bands have been discovered
experimentally in a number of nearly spherical nuclei in
$A = 80, 110, 135$, and 190 mass regions~\cite{Clark2000_ARNPS50-1,
Frauendorf2001_RMP73-463, Hubel2005_PPNP54-1,
Meng2013_FrontiersofPhysics8-55, *Meng2013_arXiv:1301.1808}.
The interpretation of such bands in terms of the shears mechanism
was firstly given in Ref.~\cite{Frauendorf1993_NPA557-259c}.
In order to distinguish this kind of rotation from the usual rotation
in well-deformed nuclei, the term ``magnetic rotation'' (MR)
was introduced~\cite{Frauendorf1994_Proceedings}.

In analogy to the antiferromagnetism in condensed matter physics, a similar
phenomenon known as ``antimagnetic rotation'' (AMR) is predicted in nuclei
by Frauendorf~\cite{Frauendorf2001_RMP73-463, Frauendorf1996_Proceedings}.
The AMR band can be explained by the ¡°two-shears-like¡± mechanism:
in some specific nearly spherical nuclei, two valence protons (neutrons)
are aligned back to back in opposite directions, nearly
perpendicular to the orientation of the total spin of the neutrons.
A rotational band can be built on such a near-spherical nucleus
since the rotational symmetry is violated by the nucleon currents.
Higher angular momenta is obtained by simultaneously
aligning the two angular momenta for the valence protons
(neutrons) toward the neutron (proton) angular momentum vector.

AMR is expected to be observed in the same mass regions
as MR~\cite{Frauendorf2001_RMP73-463}.
However, it differs from MR in two aspects.
First, there is no $M1$ transition in the AMR band since
the transverse magnetic moments of the magnetic subsystems are antialigned.
The resulting transverse magnetic moment is zero.
Second, as the antimagnetic rotor is symmetric with respect to
a rotation by $\pi$ about the total angular momentum axis,
the AMR bands consist of regular sequences of
energy levels differing in spin by $2\hbar$ and are connected
by weak $E2$ transitions reflecting the nearly spherical core.
Moreover, the phenomenon of AMR is
characterized by a decrease of the $B(E2)$ values with spin.
Since AMR was proposed~\cite{Frauendorf2001_RMP73-463},
it has been investigated both from experimental and theoretical aspects.
To date, experimental evidence of AMR has been reported in Cd isotopes
including $^{105}$Cd~\cite{Choudhury2010_PRC82-061308R},
$^{106}$Cd~\cite{Simons2003_PRL91-162501},
$^{108}$Cd~\cite{Simons2005_PRC72-024318, Datta2005_PRC71-041305R},
$^{110}$Cd~\cite{Roy2011_PLB694-322}.
Most recently, two AMR bands in a single nucleus are firstly
observed in ${}^{107}$Cd~\cite{Choudhury2013_PRC87-034304}.
In addition, the occurrence of this phenomenon still needs
further investigation by lifetime measurements in
$^{109}$Cd~\cite{Chiara2000_PRC61-034318},
$^{100}$Pd~\cite{Zhu2001_PRC64-041302R},
$^{144}$Dy~\cite{Sugawara2009_PRC79-064321},
$^{101}$Pd~\cite{Sugawara2012_PRC86-034326},
and
$^{112}$In~\cite{Li2012_PRC86-057305}.

Theoretically, AMR has been discussed by simple
geometry in the classical particle rotor model~
\cite{Clark2000_ARNPS50-1},
and the tilted axis cranking (TAC) model~
\cite{Frauendorf2000_NPA677-115, Peng2008_PRC78-024313,
Zhao2011_PLB699-181}.
Based on the TAC model, many applications have been carried out
in the framework of microscopic-macroscopic model~
\cite{Zhu2001_PRC64-041302R, Simons2003_PRL91-162501, Simons2005_PRC72-024318},
and pairing plus quadrupole model~
\cite{Chiara2000_PRC61-034318, Frauendorf2001_RMP73-463}.
Most recently, the TAC model based on the covariant density functional theory
is used to investigate the AMR
~\cite{Zhao2011_PRL107-122501, Zhao2012_PRC85-054310, Liu2012_SSPMA55-2420}
with the point coupling effective interaction PC-PK1~\cite{Zhao2010_PRC82-054319},
for its review see Ref.~\cite{Meng2013_FrontiersofPhysics8-55}.
The quality of the cranking approximation for
principal-axis cranking~\cite{Meng1993_APS42-368},
tilted-axis cranking~\cite{Frauendorf1996_ZPA356-263},
and aplanar tilted-axis cranking~\cite{Frauendorf1997_NPA617-131}
has been discussed and tested within the particle rotor model.

In the present work, the cranked shell model (CSM) with the
pairing correlations treated by a particle-number conserving (PNC)
method~\cite{Zeng1983_NPA405-1, Zeng1994_PRC50-1388} is used
to investigate the AMR bands in $^{105,106}$Cd.
The PNC-CSM method is proposed to treat properly  the
pairing correlations and the
blocking effects, and it has been applied successfully for
describing the odd-even differences in moments of inertia
(MOI's)~\cite{Zeng1994_PRC50-746},
the nonadditivity in MOI's~\cite{Liu2002_PRC66-067301},
the microscopic mechanism of identical bands
~\cite{Zeng2001_PRC63-024305, Liu2002_PRC66-024320},
the non-existence of nuclear pairing phase transition
~\cite{Wu2011_PRC83-034323}, etc.
The high-spin states and high-$K$ isomers in the rare-earth,
the actinide region and superheavy nuclei are also well described
in the PNC-CSM scheme~\cite{Liu2004_NPA735-77, Zhang2009_NPA816-19,
Zhang2009_PRC80-034313, He2005_NPA760-263, Zhang2011_PRC83-011304R,
Zhang2012_PRC85-014324, Zhang2012_arxiv1208.1156v1}.
In contrary to the conventional Bardeen-Cooper-Schrieffer (BCS) or
Hartree-Fock-Bogolyubov (HFB) approach, the Hamiltonian is solved directly
in a truncated Fock-space in the PNC method~\cite{Wu1989_PRC39-666}.
Therefore, the particle-number is conserved and the Pauli blocking effects
are taken into account exactly.
The PNC scheme has been used both in relativistic and nonrelativistic
mean field models~\cite{Meng2006_FPC1-38, Pillet2002_NPA697-141}
in which the single-particle levels are calculated from the self-consistent
mean-field potentials instead of the Nilsson potential.

\section{\label{Sec:PNC-CSM}Theoretical framework}

The cranked shell model Hamiltonian of an axially symmetric
nucleus in the rotating frame can be written as
\begin{eqnarray}
 H_\mathrm{CSM}
 & = &
 H_0 + H_\mathrm{P}
 = H_{\rm Nil}-\omega J_x + H_\mathrm{P}
 \ ,
 \label{eq:H_CSM}
\end{eqnarray}
where $H_{\rm Nil}$ is the Nilsson Hamiltonian, $-\omega J_x$ is the
Coriolis interaction with the cranking frequency $\omega$ about the
$x$ axis (perpendicular to the nuclear symmetry $z$ axis).
$H_{\rm P}$ is the pairing interaction,
\begin{eqnarray}
 H_{\rm P}
 & = &
  -G \sum_{\xi\eta} a^\dag_{\xi} a^\dag_{\bar{\xi}}
                        a_{\bar{\eta}} a_{\eta}
  \ ,
\end{eqnarray}
where $\bar{\xi}$ ($\bar{\eta}$) labels the time-reversed state of a
Nilsson state $\xi$ ($\eta$),
$G$ is the effective strength of monopole pairing interaction.

Instead of the usual single-particle level truncation in conventional
shell-model calculations, a cranked many-particle configuration
(CMPC) truncation (Fock space truncation) is adopted~\cite{Zeng1994_PRC50-1388,
Molique1997_PRC56-1795}.
By diagonalizing the $H_\mathrm{CSM}$ in a sufficiently
large CMPC space, sufficiently accurate solutions for low-lying excited
eigenstates of $H_\mathrm{CSM}$ can be obtained.
An eigenstate of $H_\mathrm{CSM}$ can be written as
\begin{equation}
 |\Psi\rangle = \sum_{i} C_i \left| i \right\rangle
 \ ,
 \qquad (C_i \; \textrm{real}),
\end{equation}
where $| i \rangle$ is a CMPC (an eigenstate of the one-body operator $H_0$).
The expectation value of a one-body operator
$\mathcal {O} = \sum_{k=1}^N \mathscr{O}(k)$ is  thus written as
\begin{equation}
 \left\langle \Psi | \mathcal {O} | \Psi \right\rangle
 =\sum_i C_i^2 \left\langle i | \mathcal {O} | i \right\rangle
 +2\sum_{i<j} C_i C_j \left\langle i | \mathcal {O} | j \right\rangle \ .
\end{equation}
As $\mathcal {O}$ is a one-body operator, the matrix element
$\langle i | \mathcal {O} | j \rangle$ for $i\neq j$ is nonzero only when
$|i\rangle$ and $|j\rangle$ differ by one particle
occupation~\cite{Zeng1994_PRC50-1388}.
After a certain permutation of
creation operators, $|i\rangle$ and $|j\rangle$ can be recast into
\begin{equation}
 | i \rangle = (-1)^{M_{i\mu}} | \mu \cdots \rangle \ , \qquad
 | j \rangle = (-1)^{M_{j\nu}} | \nu \cdots \rangle \ ,
\end{equation}
where the ellipsis ``$\cdots$''
stands for the same particle occupation and
$(-1)^{M_{i\mu(\nu)}}=\pm1$ according to whether the permutation is
even or odd. Therefore, the  expectation value of $\mathcal {O}$
can be separated into the diagonal and the off-diagonal parts
\begin{eqnarray}
 \mathcal {O}
  &=& \left\langle \Psi | \mathcal {O} | \Psi \right\rangle =
  \left( \sum_{\mu} \mathscr{O}(\mu) + 2\sum_{\mu<\nu} \mathscr{O}(\mu\nu) \right) \ ,\\
 \mathscr{O}(\mu)
 &=& \langle \mu | \mathscr{O} | \mu \rangle n_{\mu}  \ , \label{eq:j1d} \\
 \mathscr{O}(\mu\nu)
 &=&\langle \mu | \mathscr{O} | \nu \rangle
  \sum_{i<j} (-1)^{M_{i\mu}+M_{j\nu}} C_{i} C_{j} \ ,
  \label{eq:j1od}
\end{eqnarray}
where $n_{\mu} = \sum_{i} |C_{i}|^{2} P_{i\mu}$ is the
occupation probability of the cranked Nilsson orbital $|\mu\rangle$
and $P_{i\mu}=1$ (0) if $|\mu\rangle$ is occupied (empty) in $|i\rangle$.

The kinematic moment of inertia $J^{(1)}$ of $|\Psi\rangle$ can be written as
\begin{eqnarray}
 J^{(1)} = \frac{1}{\omega} \left\langle \Psi | J_x | \Psi \right\rangle \ .
\end{eqnarray}
The $B(E2)$ transition probabilities can be derived
in the semiclassical approximation as
\begin{equation}
B(E2) = \frac{3}{8}
{\left\langle \Psi | Q_{20}^{\rm p} | \Psi \right\rangle}^2 \ ,
\end{equation}
where $Q_{20}^{\rm p}$ correspond to
the quadrupole moments of protons and
\begin{equation}
Q_{20} = \sqrt{\frac{5}{16\pi}} (3z^2-r^2) = r^2 Y_{20}.
\end{equation}

\section{\label{Sec:Results}Results and discussions}

The Nilsson parameters ($\kappa$ and $\mu$) of $^{105,106}$Cd are taken
from the Lund systematics~\cite{Nilsson1969_NPA131-1}.
The quadurpole deformation parameters are taken as
$\varepsilon_2=0.12$ and $\varepsilon_2=0.14$ for
$^{105}$Cd and $^{106}$Cd, respectively.
These values are close to those used in the TAC calculations based on
the microscopic-macroscopic model or  the covariant density functional
theory~\cite{Simons2003_PRL91-162501, Zhao2011_PRL107-122501}.
The valence single-particle space in this work is constructed
in the major shells from $N=0$ to $N=5$ both for protons and neutrons,
so there is no effective charge involved in the calculation of the $B(E2)$ values.
The effective pairing strengths can, in principle,
be determined by the odd-even differences in nuclear masses and the MOI's,
and are connected with the dimension of the truncated CMPC space.
The dimensions of the CMPC space are about 1000 both for protons and neutrons.
The corresponding effective pairing strengths used in this work are
$G_p$ = 0.45~MeV and $G_{n}$ = 0.80~MeV for $^{105}$Cd,
$G_p$ = 0.45~MeV and $G_{n}$ = 0.45~MeV for $^{106}$Cd.
The data show that the MOI's for the AMR band in $^{105}$Cd
are smaller than those in $^{106}$Cd.
Therefore, a larger effective neutron pairing strength for $^{105}$Cd is adopted.
A larger CMPC space with renormalized pairing strengths gives essentially the same results.
In addition, the stability of the PNC-CSM calculation results against the change
of the dimension of the CMPC space has been investigated in Refs.
~\cite{Zeng1994_PRC50-1388, Liu2002_PRC66-024320, Zhang2012_PRC85-014324}.
In the present calculations, almost all the important CMPC's
(with the corresponding weights larger than $0.1\%$) are taken into account,
so the solutions to the low-lying excited states are accurate enough.

\begin{figure}[h]
\includegraphics[scale=0.8]{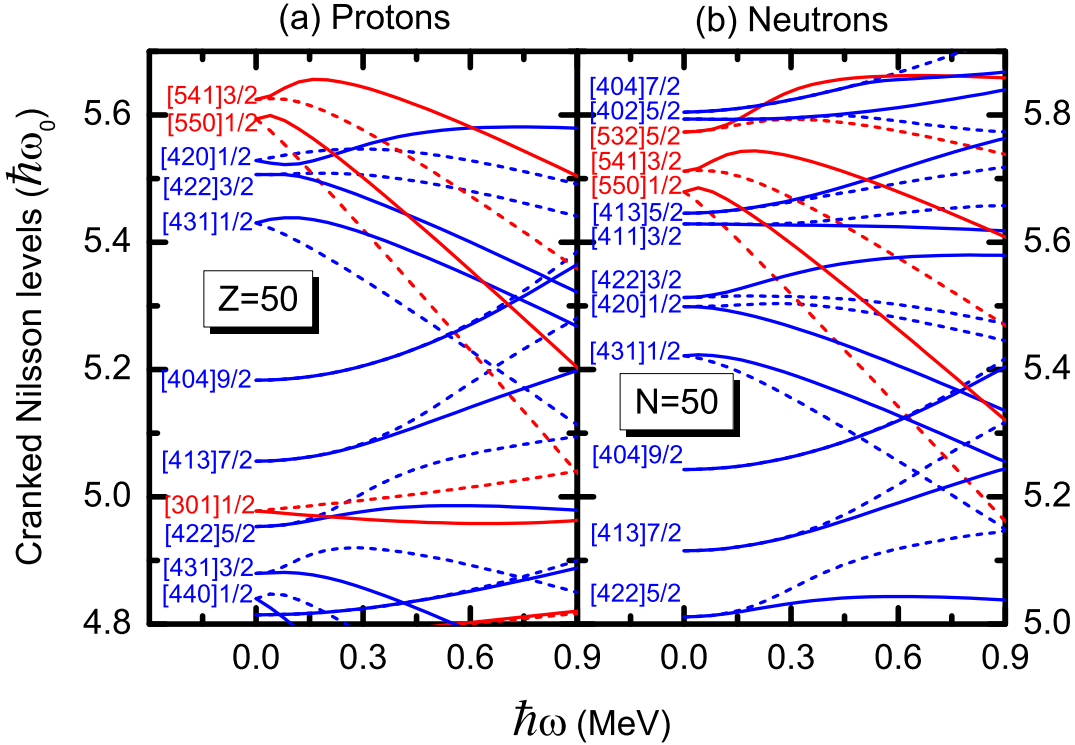}
\caption{\label{fig:Nilsson} (Color online).
The cranked Nilsson levels near the Fermi surface of $^{106}$Cd
for (a) protons and (b) neutrons.
The positive (negative) parity levels are denoted by blue (red) lines.
The signature $\alpha=+1/2$ ($\alpha=-1/2$) levels
are denoted by solid (dotted) lines.
The deformation parameter $\varepsilon_2 = 0.14$.}
\end{figure}

The cranked Nilsson levels near the Fermi surface of $^{106}$Cd
for (a) protons and (b) neutrons are given in Fig.~\ref{fig:Nilsson}.
The positive (negative) parity levels are denoted by blue (red) lines.
The signature $\alpha=+1/2$ ($\alpha=-1/2$)
levels are denoted by solid (dotted) lines.
Because the Nilsson levels of $^{105}$Cd are very similar
with those of the $^{106}$Cd, we do not show them here.
It can be seen from Fig.~\ref{fig:Nilsson} that the two proton
holes for $^{105,106}$Cd are $\pi 9/2^+[404] (\pi g_{9/2})$.
The data show that the AMR bands in
$^{105}$Cd~\cite{Choudhury2010_PRC82-061308R} and
$^{106}$Cd~\cite{Simons2003_PRL91-162501}
are the lowest lying negative and positive parity band, respectively.
The lowest lying negative parity band for $^{105}$Cd in our calculation
is $ \nu 1/2^-[550] (h_{11/2})$ %with the signature $\alpha = -1/2$
and the lowest lying positive parity band for $^{106}$Cd is the yrast band.
Therefore, in the following investigation,
we do nonadiabatic calculations for the $\nu 1/2^-[550]$ %($\alpha = -1/2$)
band in $^{105}$Cd and the yrast band in $^{106}$Cd.

\begin{figure}[h]
\includegraphics[scale=0.5]{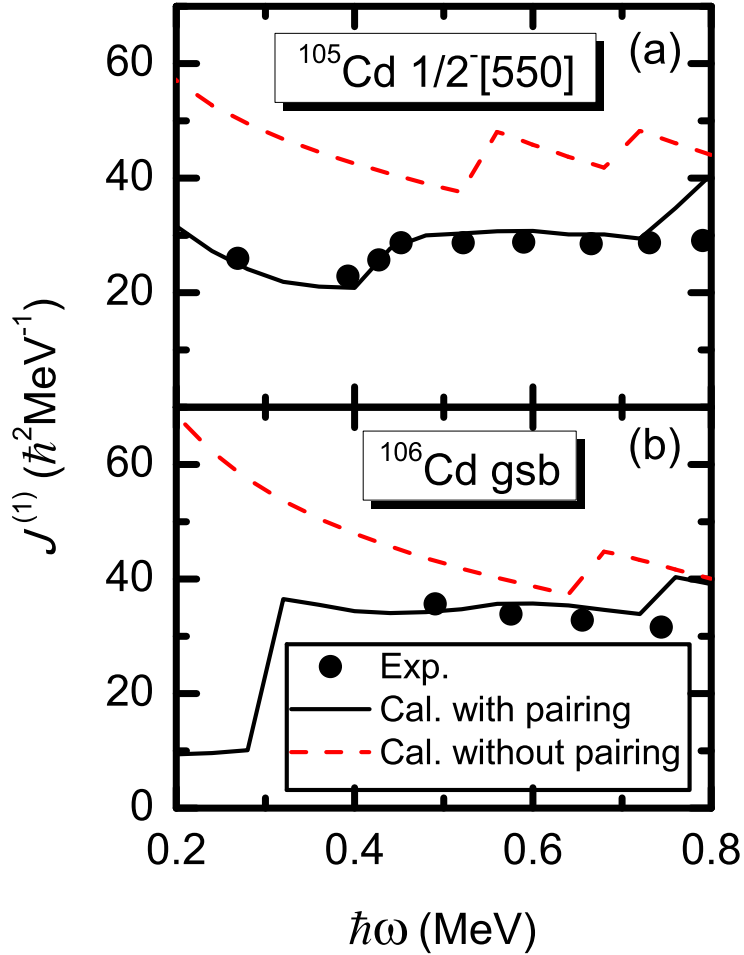}
\caption{\label{fig:MOI} (Color online).
The experimental (solid circles) and calculated kinematic MOI's
$J^{(1)}$ with (solid black lines) and without (dashed red lines)
pairing correlations for (a) $^{105}$Cd and (b) $^{106}$Cd.}
\end{figure}

Figure~\ref{fig:MOI} shows the experimental (solid circles)
and calculated kinematic MOI's $J^{(1)}$ with (solid black lines)
and without (dashed red lines) pairing correlations
for $^{105}$Cd (upper panel) and $^{106}$Cd (lower paner).
The pairing interaction is very important in reproducing
the experimental MOI's, especially the upbending.
It can be seen that the MOI's of $^{105,106}$Cd are
overestimated when the pairing interaction is switched off,
while they are well reproduced after considering the pairing correlations.
The first backbending in $^{105}$Cd at $\hbar\omega \approx 0.4$~MeV
is caused by aligning one neutron pair $\nu g_{7/2}$.
The configuration after backbending in $^{105}$Cd is thus
$\nu h_{11/2} (g_{7/2})^2$ coupled to a pair of $\pi g_{9/2}$ proton holes,
which is consistent with the previous calculations~
\cite{Choudhury2010_PRC82-061308R, Zhao2011_PRL107-122501}.
The first backbending in $^{106}$Cd at $\hbar\omega \approx 0.3$~MeV
is caused by one pair of neutrons jumping from $\nu g_{7/2}$ to $\nu h_{11/2}$.

\begin{figure}[h]
\includegraphics[scale=0.5]{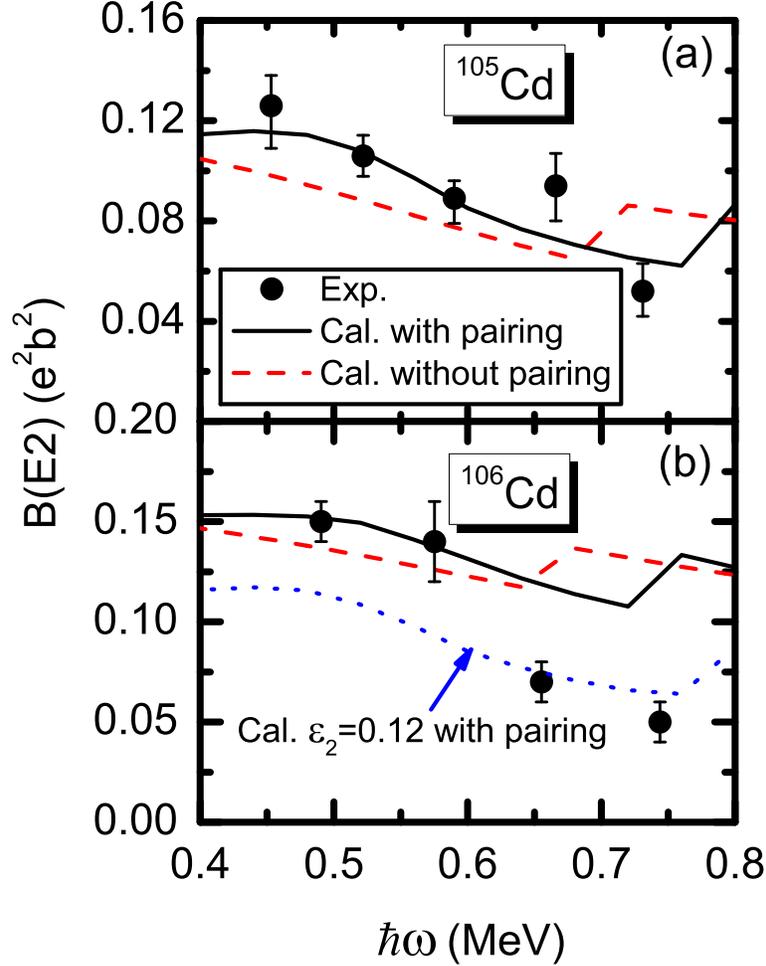}
\caption{\label{fig:BE2} (Color online).
The experimental (solid circles) and calculated $B(E2)$ values
with (solid black lines) and without (dashed red lines)
pairing correlations for (a) $^{105}$Cd and (b) $^{106}$Cd.
The blue dotted line in (b) is the calculated results
with a reduced deformation of $\varepsilon_2 = 0.12$
in which the pairing in considered.
The data for $^{105, 106}$Cd are taken from~
\cite{Choudhury2010_PRC82-061308R, Simons2003_PRL91-162501}.}
\end{figure}

One of the typical features of AMR is the weak $E2$ transitions
reflecting the small deformation of the core.
Moreover, the corresponding $B(E2)$ values rapidly decrease
with the angular momentum, which is connected with the
interpretation of the two-shears-like mechanism.
Figure~\ref{fig:BE2} shows the experimental (solid circles)
and calculated $B(E2)$ values with (solid black lines)
and without (dashed red lines) pairing correlations
for $^{105}$Cd (upper panel) and $^{106}$Cd (lower panel).
It can be seen that the decreasing tendency of the $B(E2)$ values
with the cranking frequency can be obtained no matter the
pairing correlation is considered or not.
However, the agreement between the data and the calculated results
is further improved by taking the pairing correlation into account,
especially for the higher rotational frequency region.
For $^{105}$Cd, with paring correlations, the expectation value of
$Q_{20}$ decreases from 0.55~eb to 0.41~eb with the rotational frequency
$\hbar\omega$ increasing from 0.45~MeV to 0.75~MeV.
The $Q_{20}$ value and the corresponding $B(E2)$ value
are reduced to about 75\% and 55\%, respectively,
which are caused by the effect of the cranking.
For $^{106}$Cd, it is difficult to describe the $B(E2)$ behavior
with a frozen deformation parameter.
This may be due to the deformation change with the rotational frequency for $^{106}$Cd.
In fact, as show in the blue dotted line in Fig.~\ref{fig:BE2}(b),
in order to reproduce the $B(E2)$ behavior from
$\hbar\omega = 0.45$~MeV to $ \hbar\omega = 0.75$~MeV,
a corresponding deformation change from $\varepsilon_2 = 0.14$ to
$\varepsilon_2 = 0.12$ is necessary.
Therefore, it can be seen that the two-shears-like mechanism
alone can provide the decrease of the $B(E2)$ values in $^{105}$Cd,
while additional reduction of the deformation is needed for $^{106}$Cd.

\begin{figure}[h]
\includegraphics[scale=0.5]{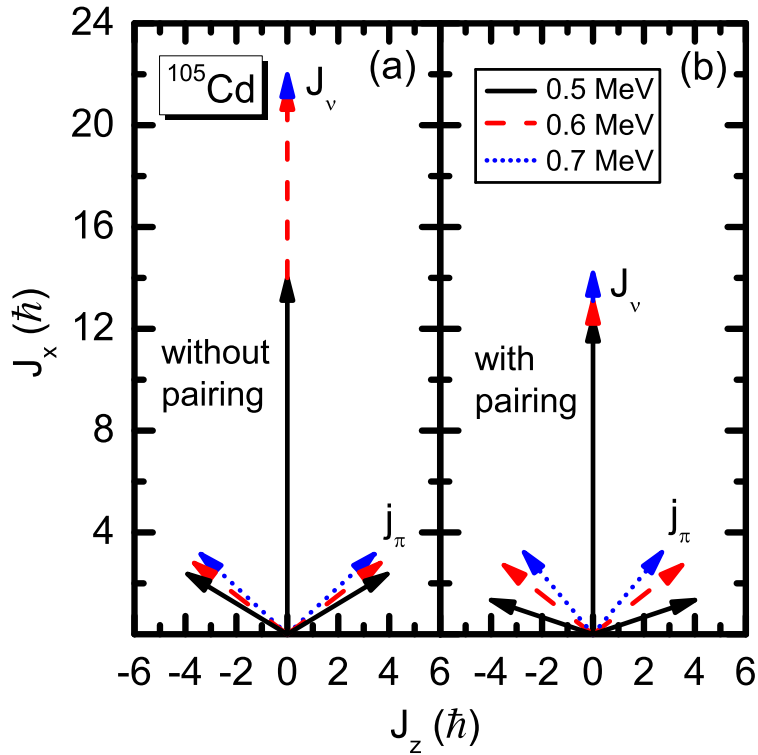}
\caption{\label{fig:JxJz} (Color online).
Angular momentum vectors of neutrons
$J_\nu$ and the two $\pi g_{9/2}$ proton holes $j_\pi$,
(a) without pairing (b) with pairing,
at rotational frequencies from 0.5 to 0.7~MeV in $^{105}$Cd.}
\end{figure}

In order to examine the two-shears-like mechanism for the
AMR band, we show the angular momentum vectors of neutrons
$J_\nu$ and the two $\pi g_{9/2}$ proton holes $j_\pi$ at rotational
frequencies from 0.5 to 0.7 MeV in $^{105}$Cd in Fig.~\ref{fig:JxJz}.
It should be mentioned that, in the principal axis cranking model,
the expectation value of $J_z$ vanishes due to the conservation of signature.
In the present AMR bands, the two proton holes in both
$^{105}$Cd and $^{106}$Cd are paired.
This means that the total angular momentum projection $K$
of these two proton holes should always be zero.
The angular momenta of the two proton holes could,
in principle, be extracted exactly from the TAC calculation.
Here, we calculate $J_z$ approximately in the following way
according to Ref.~\cite{Frauendorf1996_ZPA356-263}
\begin{equation}
J_z = \sqrt{\langle \Psi | J_z^2 | \Psi \rangle} \ .
\end{equation}
This method has been proved to be a good approximation
by comparing the principal axis cranking with
the particle rotor model in Ref.~\cite{Frauendorf1996_ZPA356-263}.
It can be seen from Fig.~\ref{fig:JxJz} that the two proton angular momentum
vectors $j_\pi$ are pointing opposite to each other and are nearly
perpendicular to the vector $J_\nu$ at $\hbar\omega = 0.5$~MeV.
The abrupt increasing of neutron angular momentum alignment
from $\hbar\omega =0.5$ to 0.6~MeV in Fig.~\ref{fig:JxJz}(a)
is due to level crossing.
After considering the nuclear superfluidity, the level crossing is
delayed and the neutron angular momentum alignment increases gradually,
which is consistent with the data.
With increasing cranking frequency the gradual alignment of
the vectors $j_\pi$ of the two $\pi g_{9/2}$ proton holes
toward the vector $J_\nu$ generates angular momentum while the
direction of the total angular momentum stays unchanged.
This leads to the closing of the two shears.
The two-shears-like mechanism can thus be clearly seen.
It should be noted that the closing of the two proton hole
angular momenta becomes more obvious when the pairing
correlation is taken into account.
This indicates the important role played by the the nuclear superfluidity in AMR.

\section{\label{Sec:Summary}Summary}

In summary, antimagnetic rotation bands in $^{105}$Cd and $^{106}$Cd are
investigated by the cranked shell model with the pairing correlations treated by a
particle-number conserving method and the blocking effects taken into account exactly.
The experimental moments of inertia in $^{105}$Cd and $^{106}$Cd
are excellently reproduced with the proper treatment of the nuclear superfluidity.
The reduced $B(E2)$ transition depends on the deformation rather than the superfluidity.
The calculated $B(E2)$ values in $^{105}$Cd are in good agreement with the data.
In order to reproduce the $B(E2)$ values in $^{106}$Cd,
a corresponding deformation change is necessary.
The two-shears-like mechanism for the antimagnetic rotation is investigated
and its sensitive dependence on the nuclear superfluidity is revealed.

\begin{acknowledgements}

This work has been supported by
National Key Basic Research Program of China (Grant No. 2013CB834400),
National Natural Science Foundation of China
(Grant No. 11121403, No. 11175002, No. 11275248, No. 11175252,
No. 11120101005, and No. 11211120152),
Knowledge Innovation Project of Chinese Academy of Sciences
(Grant No. KJCX2-EW-N01 and No. KJCX2-YW-N32),
the Research Fund for the Doctoral Program of Higher Education
under Grant No. 20110001110087,
and the China Postdoctoral Science Foundation under Grant No. 2012M520101.
The results described in this paper are obtained on the
ScGrid of Supercomputing Center,
Computer Network Information Center of Chinese Academy of Sciences.

\end{acknowledgements}

%\end{CJK*}

%\bibliography{../../../../Refecences/Papers}

\begin{thebibliography}{48}%
\makeatletter
\providecommand \@ifxundefined [1]{%
 \@ifx{#1\undefined}
}%
\providecommand \@ifnum [1]{%
 \ifnum #1\expandafter \@firstoftwo
 \else \expandafter \@secondoftwo
 \fi
}%
\providecommand \@ifx [1]{%
 \ifx #1\expandafter \@firstoftwo
 \else \expandafter \@secondoftwo
 \fi
}%
\providecommand \natexlab [1]{#1}%
\providecommand \enquote  [1]{``#1''}%
\providecommand \bibnamefont  [1]{#1}%
\providecommand \bibfnamefont [1]{#1}%
\providecommand \citenamefont [1]{#1}%
\providecommand \href@noop [0]{\@secondoftwo}%
\providecommand \href [0]{\begingroup \@sanitize@url \@href}%
\providecommand \@href[1]{\@@startlink{#1}\@@href}%
\providecommand \@@href[1]{\endgroup#1\@@endlink}%
\providecommand \@sanitize@url [0]{\catcode `\\12\catcode `\$12\catcode
  `\&12\catcode `\#12\catcode `\^12\catcode `\_12\catcode `\%12\relax}%
\providecommand \@@startlink[1]{}%
\providecommand \@@endlink[0]{}%
\providecommand \url  [0]{\begingroup\@sanitize@url \@url }%
\providecommand \@url [1]{\endgroup\@href {#1}{\urlprefix }}%
\providecommand \urlprefix  [0]{URL }%
\providecommand \Eprint [0]{\href }%
\providecommand \doibase [0]{http://dx.doi.org/}%
\providecommand \selectlanguage [0]{\@gobble}%
\providecommand \bibinfo  [0]{\@secondoftwo}%
\providecommand \bibfield  [0]{\@secondoftwo}%
\providecommand \translation [1]{[#1]}%
\providecommand \BibitemOpen [0]{}%
\providecommand \bibitemStop [0]{}%
\providecommand \bibitemNoStop [0]{.\EOS\space}%
\providecommand \EOS [0]{\spacefactor3000\relax}%
\providecommand \BibitemShut  [1]{\csname bibitem#1\endcsname}%
\let\auto@bib@innerbib\@empty
%</preamble>
\bibitem [{\citenamefont {H\"ubel}(2005)}]{Hubel2005_PPNP54-1}%
  \BibitemOpen
  \bibfield  {author} {\bibinfo {author} {\bibfnamefont {H.}~\bibnamefont
  {H\"ubel}},\ }\href {\doibase 10.1016/j.ppnp.2004.06.002} {\bibfield
  {journal} {\bibinfo  {journal} {Prog. Part. Nucl. Phys.}\ }\textbf {\bibinfo
  {volume} {54}},\ \bibinfo {pages} {1} (\bibinfo {year} {2005})}\BibitemShut
  {NoStop}%
\bibitem [{\citenamefont {Clark}\ and\ \citenamefont
  {Macchiavelli}(2000)}]{Clark2000_ARNPS50-1}%
  \BibitemOpen
  \bibfield  {author} {\bibinfo {author} {\bibfnamefont {R.~M.}\ \bibnamefont
  {Clark}}\ and\ \bibinfo {author} {\bibfnamefont {A.~O.}\ \bibnamefont
  {Macchiavelli}},\ }\href {\doibase 10.1146/annurev.nucl.50.1.1} {\bibfield
  {journal} {\bibinfo  {journal} {Annu. Rev. Nucl. Part. Sci.}\ }\textbf
  {\bibinfo {volume} {50}},\ \bibinfo {pages} {1} (\bibinfo {year}
  {2000})}\BibitemShut {NoStop}%
\bibitem [{\citenamefont {Frauendorf}(2001)}]{Frauendorf2001_RMP73-463}%
  \BibitemOpen
  \bibfield  {author} {\bibinfo {author} {\bibfnamefont {S.}~\bibnamefont
  {Frauendorf}},\ }\href {\doibase 10.1103/RevModPhys.73.463} {\bibfield
  {journal} {\bibinfo  {journal} {Rev. Mod. Phys.}\ }\textbf {\bibinfo {volume}
  {73}},\ \bibinfo {pages} {463} (\bibinfo {year} {2001})}\BibitemShut
  {NoStop}%
\bibitem [{\citenamefont {Meng}\ \emph {et~al.}(2013)\citenamefont {Meng},
  \citenamefont {Peng}, \citenamefont {Zhang},\ and\ \citenamefont
  {Zhao}}]{Meng2013_FrontiersofPhysics8-55}%
  \BibitemOpen
  \bibfield  {author} {\bibinfo {author} {\bibfnamefont {J.}~\bibnamefont
  {Meng}}, \bibinfo {author} {\bibfnamefont {J.}~\bibnamefont {Peng}}, \bibinfo
  {author} {\bibfnamefont {S.-Q.}\ \bibnamefont {Zhang}}, \ and\ \bibinfo
  {author} {\bibfnamefont {P.-W.}\ \bibnamefont {Zhao}},\ }\href {\doibase
  10.1007/s11467-013-0287-y} {\bibfield  {journal} {\bibinfo  {journal}
  {Frontiers of Physics}\ }\textbf {\bibinfo {volume} {8}},\ \bibinfo {pages}
  {55} (\bibinfo {year} {2013})}\BibitemShut {NoStop}%
\bibitem [{Men()}]{Meng2013_arXiv:1301.1808}%
  \BibitemOpen
  \href {http://arxiv.org/abs/1301.1808} {\bibinfo  {journal} {arXiv:1301.1808
  [nucl-th]}\ }\BibitemShut {NoStop}%
\bibitem [{\citenamefont {Frauendorf}(1993)}]{Frauendorf1993_NPA557-259c}%
  \BibitemOpen
\bibfield  {journal} {  }\bibfield  {author} {\bibinfo {author} {\bibfnamefont
  {S.}~\bibnamefont {Frauendorf}},\ }\href {\doibase
  10.1016/0375-9474(93)90546-A} {\bibfield  {journal} {\bibinfo  {journal}
  {Nucl. Phys. A}\ }\textbf {\bibinfo {volume} {557}},\ \bibinfo {pages} {259c}
  (\bibinfo {year} {1993})}\BibitemShut {NoStop}%
\bibitem [{\citenamefont {Frauendorf}\ \emph {et~al.}(1994)\citenamefont
  {Frauendorf}, \citenamefont {Meng},\ and\ \citenamefont
  {Reif}}]{Frauendorf1994_Proceedings}%
  \BibitemOpen
  \bibfield  {author} {\bibinfo {author} {\bibfnamefont {S.}~\bibnamefont
  {Frauendorf}}, \bibinfo {author} {\bibfnamefont {J.}~\bibnamefont {Meng}}, \
  and\ \bibinfo {author} {\bibfnamefont {J.}~\bibnamefont {Reif}},\ }in\
  \href@noop {} {\emph {\bibinfo {booktitle} {Proceedings of the Conference on
  Physics From Large $\gamma$-Ray Detector Arrays}}},\ Vol.\ \bibinfo {volume}
  {II of Report LBL35687},\ \bibinfo {editor} {edited by\ \bibinfo {editor}
  {\bibfnamefont {M.~A.}\ \bibnamefont {Deleplanque}}}\ (\bibinfo  {publisher}
  {Univ. of California, Berkeley},\ \bibinfo {year} {1994})\ p.~\bibinfo
  {pages} {52}\BibitemShut {NoStop}%
\bibitem [{\citenamefont {Frauendorf}(1996)}]{Frauendorf1996_Proceedings}%
  \BibitemOpen
  \bibfield  {author} {\bibinfo {author} {\bibfnamefont {S.}~\bibnamefont
  {Frauendorf}},\ }in\ \href@noop {} {\emph {\bibinfo {booktitle} {Proceedings
  of the Workshop on Gammasphere Physics, Berkeley, 1995}}},\ \bibinfo {editor}
  {edited by\ \bibinfo {editor} {\bibfnamefont {M.~A.}\ \bibnamefont
  {Deleplanque}}, \bibinfo {editor} {\bibfnamefont {I.~Y.}\ \bibnamefont
  {Lee}}, \ and\ \bibinfo {editor} {\bibfnamefont {A.~O.}\ \bibnamefont
  {Macchiavelli}}}\ (\bibinfo  {publisher} {World Scientific, Singapore},\
  \bibinfo {year} {1996})\ p.\ \bibinfo {pages} {272}\BibitemShut {NoStop}%
\bibitem [{\citenamefont {Choudhury}\ \emph {et~al.}(2010)\citenamefont
  {Choudhury}, \citenamefont {Jain}, \citenamefont {Patial}, \citenamefont
  {Gupta}, \citenamefont {Arumugam}, \citenamefont {Dhal}, \citenamefont
  {Sinha}, \citenamefont {Chaturvedi}, \citenamefont {Joshi}, \citenamefont
  {Trivedi}, \citenamefont {Palit}, \citenamefont {Kumar}, \citenamefont
  {Garg}, \citenamefont {Mandal}, \citenamefont {Negi}, \citenamefont
  {Mohanto}, \citenamefont {Muralithar}, \citenamefont {Singh}, \citenamefont
  {Madhavan}, \citenamefont {Bhowmik},\ and\ \citenamefont
  {Pancholi}}]{Choudhury2010_PRC82-061308R}%
  \BibitemOpen
  \bibfield  {author} {\bibinfo {author} {\bibfnamefont {D.}~\bibnamefont
  {Choudhury}}, \bibinfo {author} {\bibfnamefont {A.~K.}\ \bibnamefont {Jain}},
  \bibinfo {author} {\bibfnamefont {M.}~\bibnamefont {Patial}}, \bibinfo
  {author} {\bibfnamefont {N.}~\bibnamefont {Gupta}}, \bibinfo {author}
  {\bibfnamefont {P.}~\bibnamefont {Arumugam}}, \bibinfo {author}
  {\bibfnamefont {A.}~\bibnamefont {Dhal}}, \bibinfo {author} {\bibfnamefont
  {R.~K.}\ \bibnamefont {Sinha}}, \bibinfo {author} {\bibfnamefont
  {L.}~\bibnamefont {Chaturvedi}}, \bibinfo {author} {\bibfnamefont {P.~K.}\
  \bibnamefont {Joshi}}, \bibinfo {author} {\bibfnamefont {T.}~\bibnamefont
  {Trivedi}}, \bibinfo {author} {\bibfnamefont {R.}~\bibnamefont {Palit}},
  \bibinfo {author} {\bibfnamefont {S.}~\bibnamefont {Kumar}}, \bibinfo
  {author} {\bibfnamefont {R.}~\bibnamefont {Garg}}, \bibinfo {author}
  {\bibfnamefont {S.}~\bibnamefont {Mandal}}, \bibinfo {author} {\bibfnamefont
  {D.}~\bibnamefont {Negi}}, \bibinfo {author} {\bibfnamefont {G.}~\bibnamefont
  {Mohanto}}, \bibinfo {author} {\bibfnamefont {S.}~\bibnamefont {Muralithar}},
  \bibinfo {author} {\bibfnamefont {R.~P.}\ \bibnamefont {Singh}}, \bibinfo
  {author} {\bibfnamefont {N.}~\bibnamefont {Madhavan}}, \bibinfo {author}
  {\bibfnamefont {R.~K.}\ \bibnamefont {Bhowmik}}, \ and\ \bibinfo {author}
  {\bibfnamefont {S.~C.}\ \bibnamefont {Pancholi}},\ }\href {\doibase
  10.1103/PhysRevC.82.061308} {\bibfield  {journal} {\bibinfo  {journal} {Phys.
  Rev. C}\ }\textbf {\bibinfo {volume} {82}},\ \bibinfo {pages} {061308R}
  (\bibinfo {year} {2010})}\BibitemShut {NoStop}%
\bibitem [{\citenamefont {Simons}\ \emph {et~al.}(2003)\citenamefont {Simons},
  \citenamefont {Wadsworth}, \citenamefont {Jenkins}, \citenamefont {Clark},
  \citenamefont {Cromaz}, \citenamefont {Deleplanque}, \citenamefont {Diamond},
  \citenamefont {Fallon}, \citenamefont {Lane}, \citenamefont {Lee},
  \citenamefont {Macchiavelli}, \citenamefont {Stephens}, \citenamefont
  {Svensson}, \citenamefont {Vetter}, \citenamefont {Ward},\ and\ \citenamefont
  {Frauendorf}}]{Simons2003_PRL91-162501}%
  \BibitemOpen
  \bibfield  {author} {\bibinfo {author} {\bibfnamefont {A.~J.}\ \bibnamefont
  {Simons}}, \bibinfo {author} {\bibfnamefont {R.}~\bibnamefont {Wadsworth}},
  \bibinfo {author} {\bibfnamefont {D.~G.}\ \bibnamefont {Jenkins}}, \bibinfo
  {author} {\bibfnamefont {R.~M.}\ \bibnamefont {Clark}}, \bibinfo {author}
  {\bibfnamefont {M.}~\bibnamefont {Cromaz}}, \bibinfo {author} {\bibfnamefont
  {M.~A.}\ \bibnamefont {Deleplanque}}, \bibinfo {author} {\bibfnamefont
  {R.~M.}\ \bibnamefont {Diamond}}, \bibinfo {author} {\bibfnamefont
  {P.}~\bibnamefont {Fallon}}, \bibinfo {author} {\bibfnamefont {G.~J.}\
  \bibnamefont {Lane}}, \bibinfo {author} {\bibfnamefont {I.~Y.}\ \bibnamefont
  {Lee}}, \bibinfo {author} {\bibfnamefont {A.~O.}\ \bibnamefont
  {Macchiavelli}}, \bibinfo {author} {\bibfnamefont {F.~S.}\ \bibnamefont
  {Stephens}}, \bibinfo {author} {\bibfnamefont {C.~E.}\ \bibnamefont
  {Svensson}}, \bibinfo {author} {\bibfnamefont {K.}~\bibnamefont {Vetter}},
  \bibinfo {author} {\bibfnamefont {D.}~\bibnamefont {Ward}}, \ and\ \bibinfo
  {author} {\bibfnamefont {S.}~\bibnamefont {Frauendorf}},\ }\href {\doibase
  10.1103/PhysRevLett.91.162501} {\bibfield  {journal} {\bibinfo  {journal}
  {Phys. Rev. Lett.}\ }\textbf {\bibinfo {volume} {91}},\ \bibinfo {pages}
  {162501} (\bibinfo {year} {2003})}\BibitemShut {NoStop}%
\bibitem [{\citenamefont {Simons}\ \emph {et~al.}(2005)\citenamefont {Simons},
  \citenamefont {Wadsworth}, \citenamefont {Jenkins}, \citenamefont {Clark},
  \citenamefont {Cromaz}, \citenamefont {Deleplanque}, \citenamefont {Diamond},
  \citenamefont {Fallon}, \citenamefont {Lane}, \citenamefont {Lee},
  \citenamefont {Macchiavelli}, \citenamefont {Stephens}, \citenamefont
  {Svensson}, \citenamefont {Vetter}, \citenamefont {Ward}, \citenamefont
  {Frauendorf},\ and\ \citenamefont {Gu}}]{Simons2005_PRC72-024318}%
  \BibitemOpen
  \bibfield  {author} {\bibinfo {author} {\bibfnamefont {A.~J.}\ \bibnamefont
  {Simons}}, \bibinfo {author} {\bibfnamefont {R.}~\bibnamefont {Wadsworth}},
  \bibinfo {author} {\bibfnamefont {D.~G.}\ \bibnamefont {Jenkins}}, \bibinfo
  {author} {\bibfnamefont {R.~M.}\ \bibnamefont {Clark}}, \bibinfo {author}
  {\bibfnamefont {M.}~\bibnamefont {Cromaz}}, \bibinfo {author} {\bibfnamefont
  {M.~A.}\ \bibnamefont {Deleplanque}}, \bibinfo {author} {\bibfnamefont
  {R.~M.}\ \bibnamefont {Diamond}}, \bibinfo {author} {\bibfnamefont
  {P.}~\bibnamefont {Fallon}}, \bibinfo {author} {\bibfnamefont {G.~J.}\
  \bibnamefont {Lane}}, \bibinfo {author} {\bibfnamefont {I.~Y.}\ \bibnamefont
  {Lee}}, \bibinfo {author} {\bibfnamefont {A.~O.}\ \bibnamefont
  {Macchiavelli}}, \bibinfo {author} {\bibfnamefont {F.~S.}\ \bibnamefont
  {Stephens}}, \bibinfo {author} {\bibfnamefont {C.~E.}\ \bibnamefont
  {Svensson}}, \bibinfo {author} {\bibfnamefont {K.}~\bibnamefont {Vetter}},
  \bibinfo {author} {\bibfnamefont {D.}~\bibnamefont {Ward}}, \bibinfo {author}
  {\bibfnamefont {S.}~\bibnamefont {Frauendorf}}, \ and\ \bibinfo {author}
  {\bibfnamefont {Y.}~\bibnamefont {Gu}},\ }\href {\doibase
  10.1103/PhysRevC.72.024318} {\bibfield  {journal} {\bibinfo  {journal} {Phys.
  Rev. C}\ }\textbf {\bibinfo {volume} {72}},\ \bibinfo {pages} {024318}
  (\bibinfo {year} {2005})}\BibitemShut {NoStop}%
\bibitem [{\citenamefont {Datta}\ \emph {et~al.}(2005)\citenamefont {Datta},
  \citenamefont {Chattopadhyay}, \citenamefont {Bhattacharya}, \citenamefont
  {Ghosh}, \citenamefont {Goswami}, \citenamefont {Pal}, \citenamefont
  {Sarkar}, \citenamefont {Jain}, \citenamefont {Joshi}, \citenamefont
  {Bhowmik}, \citenamefont {Kumar}, \citenamefont {Madhavan}, \citenamefont
  {Muralithar}, \citenamefont {Rao},\ and\ \citenamefont
  {Singh}}]{Datta2005_PRC71-041305R}%
  \BibitemOpen
  \bibfield  {author} {\bibinfo {author} {\bibfnamefont {P.}~\bibnamefont
  {Datta}}, \bibinfo {author} {\bibfnamefont {S.}~\bibnamefont
  {Chattopadhyay}}, \bibinfo {author} {\bibfnamefont {S.}~\bibnamefont
  {Bhattacharya}}, \bibinfo {author} {\bibfnamefont {T.~K.}\ \bibnamefont
  {Ghosh}}, \bibinfo {author} {\bibfnamefont {A.}~\bibnamefont {Goswami}},
  \bibinfo {author} {\bibfnamefont {S.}~\bibnamefont {Pal}}, \bibinfo {author}
  {\bibfnamefont {M.~S.}\ \bibnamefont {Sarkar}}, \bibinfo {author}
  {\bibfnamefont {H.~C.}\ \bibnamefont {Jain}}, \bibinfo {author}
  {\bibfnamefont {P.~K.}\ \bibnamefont {Joshi}}, \bibinfo {author}
  {\bibfnamefont {R.~K.}\ \bibnamefont {Bhowmik}}, \bibinfo {author}
  {\bibfnamefont {R.}~\bibnamefont {Kumar}}, \bibinfo {author} {\bibfnamefont
  {N.}~\bibnamefont {Madhavan}}, \bibinfo {author} {\bibfnamefont
  {S.}~\bibnamefont {Muralithar}}, \bibinfo {author} {\bibfnamefont {P.~V.~M.}\
  \bibnamefont {Rao}}, \ and\ \bibinfo {author} {\bibfnamefont {R.~P.}\
  \bibnamefont {Singh}},\ }\href {\doibase 10.1103/PhysRevC.71.041305}
  {\bibfield  {journal} {\bibinfo  {journal} {Phys. Rev. C}\ }\textbf {\bibinfo
  {volume} {71}},\ \bibinfo {pages} {041305} (\bibinfo {year}
  {2005})}\BibitemShut {NoStop}%
\bibitem [{\citenamefont {Roy}\ \emph {et~al.}(2011)\citenamefont {Roy},
  \citenamefont {Chattopadhyay}, \citenamefont {Datta}, \citenamefont {Pal},
  \citenamefont {Bhattacharya}, \citenamefont {Bhowmik}, \citenamefont
  {Goswami}, \citenamefont {Jain}, \citenamefont {Kumar}, \citenamefont
  {Muralithar}, \citenamefont {Negi}, \citenamefont {Palit},\ and\
  \citenamefont {Singh}}]{Roy2011_PLB694-322}%
  \BibitemOpen
  \bibfield  {author} {\bibinfo {author} {\bibfnamefont {S.}~\bibnamefont
  {Roy}}, \bibinfo {author} {\bibfnamefont {S.}~\bibnamefont {Chattopadhyay}},
  \bibinfo {author} {\bibfnamefont {P.}~\bibnamefont {Datta}}, \bibinfo
  {author} {\bibfnamefont {S.}~\bibnamefont {Pal}}, \bibinfo {author}
  {\bibfnamefont {S.}~\bibnamefont {Bhattacharya}}, \bibinfo {author}
  {\bibfnamefont {R.}~\bibnamefont {Bhowmik}}, \bibinfo {author} {\bibfnamefont
  {A.}~\bibnamefont {Goswami}}, \bibinfo {author} {\bibfnamefont
  {H.}~\bibnamefont {Jain}}, \bibinfo {author} {\bibfnamefont {R.}~\bibnamefont
  {Kumar}}, \bibinfo {author} {\bibfnamefont {S.}~\bibnamefont {Muralithar}},
  \bibinfo {author} {\bibfnamefont {D.}~\bibnamefont {Negi}}, \bibinfo {author}
  {\bibfnamefont {R.}~\bibnamefont {Palit}}, \ and\ \bibinfo {author}
  {\bibfnamefont {R.}~\bibnamefont {Singh}},\ }\href {\doibase
  10.1016/j.physletb.2010.10.018} {\bibfield  {journal} {\bibinfo  {journal}
  {Phys. Lett. B}\ }\textbf {\bibinfo {volume} {694}},\ \bibinfo {pages} {322}
  (\bibinfo {year} {2011})}\BibitemShut {NoStop}%
\bibitem [{\citenamefont {Choudhury}\ \emph {et~al.}(2013)\citenamefont
  {Choudhury}, \citenamefont {Jain}, \citenamefont {Kumar}, \citenamefont
  {Kumar}, \citenamefont {Singh}, \citenamefont {Singh}, \citenamefont
  {Sainath}, \citenamefont {Trivedi}, \citenamefont {Sethi}, \citenamefont
  {Saha}, \citenamefont {Jadav}, \citenamefont {Naidu}, \citenamefont {Palit},
  \citenamefont {Jain}, \citenamefont {Chaturvedi},\ and\ \citenamefont
  {Pancholi}}]{Choudhury2013_PRC87-034304}%
  \BibitemOpen
  \bibfield  {author} {\bibinfo {author} {\bibfnamefont {D.}~\bibnamefont
  {Choudhury}}, \bibinfo {author} {\bibfnamefont {A.~K.}\ \bibnamefont {Jain}},
  \bibinfo {author} {\bibfnamefont {G.~A.}\ \bibnamefont {Kumar}}, \bibinfo
  {author} {\bibfnamefont {S.}~\bibnamefont {Kumar}}, \bibinfo {author}
  {\bibfnamefont {S.}~\bibnamefont {Singh}}, \bibinfo {author} {\bibfnamefont
  {P.}~\bibnamefont {Singh}}, \bibinfo {author} {\bibfnamefont
  {M.}~\bibnamefont {Sainath}}, \bibinfo {author} {\bibfnamefont
  {T.}~\bibnamefont {Trivedi}}, \bibinfo {author} {\bibfnamefont
  {J.}~\bibnamefont {Sethi}}, \bibinfo {author} {\bibfnamefont
  {S.}~\bibnamefont {Saha}}, \bibinfo {author} {\bibfnamefont {S.~K.}\
  \bibnamefont {Jadav}}, \bibinfo {author} {\bibfnamefont {B.~S.}\ \bibnamefont
  {Naidu}}, \bibinfo {author} {\bibfnamefont {R.}~\bibnamefont {Palit}},
  \bibinfo {author} {\bibfnamefont {H.~C.}\ \bibnamefont {Jain}}, \bibinfo
  {author} {\bibfnamefont {L.}~\bibnamefont {Chaturvedi}}, \ and\ \bibinfo
  {author} {\bibfnamefont {S.~C.}\ \bibnamefont {Pancholi}},\ }\href {\doibase
  10.1103/PhysRevC.87.034304} {\bibfield  {journal} {\bibinfo  {journal} {Phys.
  Rev. C}\ }\textbf {\bibinfo {volume} {87}},\ \bibinfo {pages} {034304}
  (\bibinfo {year} {2013})}\BibitemShut {NoStop}%
\bibitem [{\citenamefont {Chiara}\ \emph {et~al.}(2000)\citenamefont {Chiara},
  \citenamefont {Asztalos}, \citenamefont {Busse}, \citenamefont {Clark},
  \citenamefont {Cromaz}, \citenamefont {Deleplanque}, \citenamefont {Diamond},
  \citenamefont {Fallon}, \citenamefont {Fossan}, \citenamefont {Jenkins},
  \citenamefont {Juutinen}, \citenamefont {Kelsall}, \citenamefont {Kr\"ucken},
  \citenamefont {Lane}, \citenamefont {Lee}, \citenamefont {Macchiavelli},
  \citenamefont {MacLeod}, \citenamefont {Schmid}, \citenamefont {Sears},
  \citenamefont {Smith}, \citenamefont {Stephens}, \citenamefont {Vetter},
  \citenamefont {Wadsworth},\ and\ \citenamefont
  {Frauendorf}}]{Chiara2000_PRC61-034318}%
  \BibitemOpen
  \bibfield  {author} {\bibinfo {author} {\bibfnamefont {C.~J.}\ \bibnamefont
  {Chiara}}, \bibinfo {author} {\bibfnamefont {S.~J.}\ \bibnamefont
  {Asztalos}}, \bibinfo {author} {\bibfnamefont {B.}~\bibnamefont {Busse}},
  \bibinfo {author} {\bibfnamefont {R.~M.}\ \bibnamefont {Clark}}, \bibinfo
  {author} {\bibfnamefont {M.}~\bibnamefont {Cromaz}}, \bibinfo {author}
  {\bibfnamefont {M.~A.}\ \bibnamefont {Deleplanque}}, \bibinfo {author}
  {\bibfnamefont {R.~M.}\ \bibnamefont {Diamond}}, \bibinfo {author}
  {\bibfnamefont {P.}~\bibnamefont {Fallon}}, \bibinfo {author} {\bibfnamefont
  {D.~B.}\ \bibnamefont {Fossan}}, \bibinfo {author} {\bibfnamefont {D.~G.}\
  \bibnamefont {Jenkins}}, \bibinfo {author} {\bibfnamefont {S.}~\bibnamefont
  {Juutinen}}, \bibinfo {author} {\bibfnamefont {N.~S.}\ \bibnamefont
  {Kelsall}}, \bibinfo {author} {\bibfnamefont {R.}~\bibnamefont {Kr\"ucken}},
  \bibinfo {author} {\bibfnamefont {G.~J.}\ \bibnamefont {Lane}}, \bibinfo
  {author} {\bibfnamefont {I.~Y.}\ \bibnamefont {Lee}}, \bibinfo {author}
  {\bibfnamefont {A.~O.}\ \bibnamefont {Macchiavelli}}, \bibinfo {author}
  {\bibfnamefont {R.~W.}\ \bibnamefont {MacLeod}}, \bibinfo {author}
  {\bibfnamefont {G.}~\bibnamefont {Schmid}}, \bibinfo {author} {\bibfnamefont
  {J.~M.}\ \bibnamefont {Sears}}, \bibinfo {author} {\bibfnamefont {J.~F.}\
  \bibnamefont {Smith}}, \bibinfo {author} {\bibfnamefont {F.~S.}\ \bibnamefont
  {Stephens}}, \bibinfo {author} {\bibfnamefont {K.}~\bibnamefont {Vetter}},
  \bibinfo {author} {\bibfnamefont {R.}~\bibnamefont {Wadsworth}}, \ and\
  \bibinfo {author} {\bibfnamefont {S.}~\bibnamefont {Frauendorf}},\ }\href
  {\doibase 10.1103/PhysRevC.61.034318} {\bibfield  {journal} {\bibinfo
  {journal} {Phys. Rev. C}\ }\textbf {\bibinfo {volume} {61}},\ \bibinfo
  {pages} {034318} (\bibinfo {year} {2000})}\BibitemShut {NoStop}%
\bibitem [{\citenamefont {Zhu}\ \emph {et~al.}(2001)\citenamefont {Zhu},
  \citenamefont {Garg}, \citenamefont {Afanasjev}, \citenamefont {Frauendorf},
  \citenamefont {Kharraja}, \citenamefont {Ghugre}, \citenamefont
  {Chintalapudi}, \citenamefont {Janssens}, \citenamefont {Carpenter},
  \citenamefont {Kondev},\ and\ \citenamefont
  {Lauritsen}}]{Zhu2001_PRC64-041302R}%
  \BibitemOpen
  \bibfield  {author} {\bibinfo {author} {\bibfnamefont {S.}~\bibnamefont
  {Zhu}}, \bibinfo {author} {\bibfnamefont {U.}~\bibnamefont {Garg}}, \bibinfo
  {author} {\bibfnamefont {A.~V.}\ \bibnamefont {Afanasjev}}, \bibinfo {author}
  {\bibfnamefont {S.}~\bibnamefont {Frauendorf}}, \bibinfo {author}
  {\bibfnamefont {B.}~\bibnamefont {Kharraja}}, \bibinfo {author}
  {\bibfnamefont {S.~S.}\ \bibnamefont {Ghugre}}, \bibinfo {author}
  {\bibfnamefont {S.~N.}\ \bibnamefont {Chintalapudi}}, \bibinfo {author}
  {\bibfnamefont {R.~V.~F.}\ \bibnamefont {Janssens}}, \bibinfo {author}
  {\bibfnamefont {M.~P.}\ \bibnamefont {Carpenter}}, \bibinfo {author}
  {\bibfnamefont {F.~G.}\ \bibnamefont {Kondev}}, \ and\ \bibinfo {author}
  {\bibfnamefont {T.}~\bibnamefont {Lauritsen}},\ }\href {\doibase
  10.1103/PhysRevC.64.041302} {\bibfield  {journal} {\bibinfo  {journal} {Phys.
  Rev. C}\ }\textbf {\bibinfo {volume} {64}},\ \bibinfo {pages} {041302R}
  (\bibinfo {year} {2001})}\BibitemShut {NoStop}%
\bibitem [{\citenamefont {Sugawara}\ \emph {et~al.}(2009)\citenamefont
  {Sugawara}, \citenamefont {Toh}, \citenamefont {Oshima}, \citenamefont
  {Koizumi}, \citenamefont {Osa}, \citenamefont {Kimura}, \citenamefont
  {Hatsukawa}, \citenamefont {Goto}, \citenamefont {Kusakari}, \citenamefont
  {Morikawa}, \citenamefont {Zhang}, \citenamefont {Zhou}, \citenamefont
  {Guo},\ and\ \citenamefont {Liu}}]{Sugawara2009_PRC79-064321}%
  \BibitemOpen
  \bibfield  {author} {\bibinfo {author} {\bibfnamefont {M.}~\bibnamefont
  {Sugawara}}, \bibinfo {author} {\bibfnamefont {Y.}~\bibnamefont {Toh}},
  \bibinfo {author} {\bibfnamefont {M.}~\bibnamefont {Oshima}}, \bibinfo
  {author} {\bibfnamefont {M.}~\bibnamefont {Koizumi}}, \bibinfo {author}
  {\bibfnamefont {A.}~\bibnamefont {Osa}}, \bibinfo {author} {\bibfnamefont
  {A.}~\bibnamefont {Kimura}}, \bibinfo {author} {\bibfnamefont
  {Y.}~\bibnamefont {Hatsukawa}}, \bibinfo {author} {\bibfnamefont
  {J.}~\bibnamefont {Goto}}, \bibinfo {author} {\bibfnamefont {H.}~\bibnamefont
  {Kusakari}}, \bibinfo {author} {\bibfnamefont {T.}~\bibnamefont {Morikawa}},
  \bibinfo {author} {\bibfnamefont {Y.~H.}\ \bibnamefont {Zhang}}, \bibinfo
  {author} {\bibfnamefont {X.~H.}\ \bibnamefont {Zhou}}, \bibinfo {author}
  {\bibfnamefont {Y.~X.}\ \bibnamefont {Guo}}, \ and\ \bibinfo {author}
  {\bibfnamefont {M.~L.}\ \bibnamefont {Liu}},\ }\href {\doibase
  10.1103/PhysRevC.79.064321} {\bibfield  {journal} {\bibinfo  {journal} {Phys.
  Rev. C}\ }\textbf {\bibinfo {volume} {79}},\ \bibinfo {pages} {064321}
  (\bibinfo {year} {2009})}\BibitemShut {NoStop}%
\bibitem [{\citenamefont {Sugawara}\ \emph {et~al.}(2012)\citenamefont
  {Sugawara}, \citenamefont {Hayakawa}, \citenamefont {Oshima}, \citenamefont
  {Toh}, \citenamefont {Osa}, \citenamefont {Matsuda}, \citenamefont {Shizuma},
  \citenamefont {Hatsukawa}, \citenamefont {Kusakari}, \citenamefont
  {Morikawa}, \citenamefont {Gan},\ and\ \citenamefont
  {Czosnyka}}]{Sugawara2012_PRC86-034326}%
  \BibitemOpen
  \bibfield  {author} {\bibinfo {author} {\bibfnamefont {M.}~\bibnamefont
  {Sugawara}}, \bibinfo {author} {\bibfnamefont {T.}~\bibnamefont {Hayakawa}},
  \bibinfo {author} {\bibfnamefont {M.}~\bibnamefont {Oshima}}, \bibinfo
  {author} {\bibfnamefont {Y.}~\bibnamefont {Toh}}, \bibinfo {author}
  {\bibfnamefont {A.}~\bibnamefont {Osa}}, \bibinfo {author} {\bibfnamefont
  {M.}~\bibnamefont {Matsuda}}, \bibinfo {author} {\bibfnamefont
  {T.}~\bibnamefont {Shizuma}}, \bibinfo {author} {\bibfnamefont
  {Y.}~\bibnamefont {Hatsukawa}}, \bibinfo {author} {\bibfnamefont
  {H.}~\bibnamefont {Kusakari}}, \bibinfo {author} {\bibfnamefont
  {T.}~\bibnamefont {Morikawa}}, \bibinfo {author} {\bibfnamefont {Z.~G.}\
  \bibnamefont {Gan}}, \ and\ \bibinfo {author} {\bibfnamefont
  {T.}~\bibnamefont {Czosnyka}},\ }\href {\doibase 10.1103/PhysRevC.86.034326}
  {\bibfield  {journal} {\bibinfo  {journal} {Phys. Rev. C}\ }\textbf {\bibinfo
  {volume} {86}},\ \bibinfo {pages} {034326} (\bibinfo {year}
  {2012})}\BibitemShut {NoStop}%
\bibitem [{\citenamefont {Li}\ \emph {et~al.}(2012)\citenamefont {Li},
  \citenamefont {Li}, \citenamefont {Lu}, \citenamefont {Ma}, \citenamefont
  {Wu}, \citenamefont {Zhu}, \citenamefont {He}, \citenamefont {Li},
  \citenamefont {Zheng}, \citenamefont {Li}, \citenamefont {Wu}, \citenamefont
  {Ma},\ and\ \citenamefont {Liu}}]{Li2012_PRC86-057305}%
  \BibitemOpen
  \bibfield  {author} {\bibinfo {author} {\bibfnamefont {X.~W.}\ \bibnamefont
  {Li}}, \bibinfo {author} {\bibfnamefont {J.}~\bibnamefont {Li}}, \bibinfo
  {author} {\bibfnamefont {J.~B.}\ \bibnamefont {Lu}}, \bibinfo {author}
  {\bibfnamefont {K.~Y.}\ \bibnamefont {Ma}}, \bibinfo {author} {\bibfnamefont
  {Y.~H.}\ \bibnamefont {Wu}}, \bibinfo {author} {\bibfnamefont {L.~H.}\
  \bibnamefont {Zhu}}, \bibinfo {author} {\bibfnamefont {C.~Y.}\ \bibnamefont
  {He}}, \bibinfo {author} {\bibfnamefont {X.~Q.}\ \bibnamefont {Li}}, \bibinfo
  {author} {\bibfnamefont {Y.}~\bibnamefont {Zheng}}, \bibinfo {author}
  {\bibfnamefont {G.~S.}\ \bibnamefont {Li}}, \bibinfo {author} {\bibfnamefont
  {X.~G.}\ \bibnamefont {Wu}}, \bibinfo {author} {\bibfnamefont {Y.~J.}\
  \bibnamefont {Ma}}, \ and\ \bibinfo {author} {\bibfnamefont {Y.~Z.}\
  \bibnamefont {Liu}},\ }\href {\doibase 10.1103/PhysRevC.86.057305} {\bibfield
   {journal} {\bibinfo  {journal} {Phys. Rev. C}\ }\textbf {\bibinfo {volume}
  {86}},\ \bibinfo {pages} {057305} (\bibinfo {year} {2012})}\BibitemShut
  {NoStop}%
\bibitem [{\citenamefont {Frauendorf}(2000)}]{Frauendorf2000_NPA677-115}%
  \BibitemOpen
  \bibfield  {author} {\bibinfo {author} {\bibfnamefont {S.}~\bibnamefont
  {Frauendorf}},\ }\href {\doibase 10.1016/S0375-9474(00)00308-0} {\bibfield
  {journal} {\bibinfo  {journal} {Nucl. Phys. A}\ }\textbf {\bibinfo {volume}
  {677}},\ \bibinfo {pages} {115} (\bibinfo {year} {2000})}\BibitemShut
  {NoStop}%
\bibitem [{\citenamefont {Peng}\ \emph {et~al.}(2008)\citenamefont {Peng},
  \citenamefont {Meng}, \citenamefont {Ring},\ and\ \citenamefont
  {Zhang}}]{Peng2008_PRC78-024313}%
  \BibitemOpen
  \bibfield  {author} {\bibinfo {author} {\bibfnamefont {J.}~\bibnamefont
  {Peng}}, \bibinfo {author} {\bibfnamefont {J.}~\bibnamefont {Meng}}, \bibinfo
  {author} {\bibfnamefont {P.}~\bibnamefont {Ring}}, \ and\ \bibinfo {author}
  {\bibfnamefont {S.~Q.}\ \bibnamefont {Zhang}},\ }\href {\doibase
  10.1103/PhysRevC.78.024313} {\bibfield  {journal} {\bibinfo  {journal} {Phys.
  Rev. C}\ }\textbf {\bibinfo {volume} {78}},\ \bibinfo {pages} {024313}
  (\bibinfo {year} {2008})}\BibitemShut {NoStop}%
\bibitem [{\citenamefont {Zhao}\ \emph
  {et~al.}(2011{\natexlab{a}})\citenamefont {Zhao}, \citenamefont {Zhang},
  \citenamefont {Peng}, \citenamefont {Liang}, \citenamefont {Ring},\ and\
  \citenamefont {Meng}}]{Zhao2011_PLB699-181}%
  \BibitemOpen
  \bibfield  {author} {\bibinfo {author} {\bibfnamefont {P.~W.}\ \bibnamefont
  {Zhao}}, \bibinfo {author} {\bibfnamefont {S.~Q.}\ \bibnamefont {Zhang}},
  \bibinfo {author} {\bibfnamefont {J.}~\bibnamefont {Peng}}, \bibinfo {author}
  {\bibfnamefont {H.~Z.}\ \bibnamefont {Liang}}, \bibinfo {author}
  {\bibfnamefont {P.}~\bibnamefont {Ring}}, \ and\ \bibinfo {author}
  {\bibfnamefont {J.}~\bibnamefont {Meng}},\ }\href {\doibase
  10.1016/j.physletb.2011.03.068} {\bibfield  {journal} {\bibinfo  {journal}
  {Phys. Lett. B}\ }\textbf {\bibinfo {volume} {699}},\ \bibinfo {pages} {181 }
  (\bibinfo {year} {2011}{\natexlab{a}})}\BibitemShut {NoStop}%
\bibitem [{\citenamefont {Zhao}\ \emph
  {et~al.}(2011{\natexlab{b}})\citenamefont {Zhao}, \citenamefont {Peng},
  \citenamefont {Liang}, \citenamefont {Ring},\ and\ \citenamefont
  {Meng}}]{Zhao2011_PRL107-122501}%
  \BibitemOpen
  \bibfield  {author} {\bibinfo {author} {\bibfnamefont {P.~W.}\ \bibnamefont
  {Zhao}}, \bibinfo {author} {\bibfnamefont {J.}~\bibnamefont {Peng}}, \bibinfo
  {author} {\bibfnamefont {H.~Z.}\ \bibnamefont {Liang}}, \bibinfo {author}
  {\bibfnamefont {P.}~\bibnamefont {Ring}}, \ and\ \bibinfo {author}
  {\bibfnamefont {J.}~\bibnamefont {Meng}},\ }\href {\doibase
  10.1103/PhysRevLett.107.122501} {\bibfield  {journal} {\bibinfo  {journal}
  {Phys. Rev. Lett.}\ }\textbf {\bibinfo {volume} {107}},\ \bibinfo {pages}
  {122501} (\bibinfo {year} {2011}{\natexlab{b}})}\BibitemShut {NoStop}%
\bibitem [{\citenamefont {Zhao}\ \emph {et~al.}(2012)\citenamefont {Zhao},
  \citenamefont {Peng}, \citenamefont {Liang}, \citenamefont {Ring},\ and\
  \citenamefont {Meng}}]{Zhao2012_PRC85-054310}%
  \BibitemOpen
  \bibfield  {author} {\bibinfo {author} {\bibfnamefont {P.~W.}\ \bibnamefont
  {Zhao}}, \bibinfo {author} {\bibfnamefont {J.}~\bibnamefont {Peng}}, \bibinfo
  {author} {\bibfnamefont {H.~Z.}\ \bibnamefont {Liang}}, \bibinfo {author}
  {\bibfnamefont {P.}~\bibnamefont {Ring}}, \ and\ \bibinfo {author}
  {\bibfnamefont {J.}~\bibnamefont {Meng}},\ }\href {\doibase
  10.1103/PhysRevC.85.054310} {\bibfield  {journal} {\bibinfo  {journal} {Phys.
  Rev. C}\ }\textbf {\bibinfo {volume} {85}},\ \bibinfo {pages} {054310}
  (\bibinfo {year} {2012})}\BibitemShut {NoStop}%
\bibitem [{\citenamefont {Liu}\ and\ \citenamefont
  {Zhao}(2012)}]{Liu2012_SSPMA55-2420}%
  \BibitemOpen
  \bibfield  {author} {\bibinfo {author} {\bibfnamefont {L.}~\bibnamefont
  {Liu}}\ and\ \bibinfo {author} {\bibfnamefont {P.}~\bibnamefont {Zhao}},\
  }\href {\doibase 10.1007/s11433-012-4906-3} {\bibfield  {journal} {\bibinfo
  {journal} {Sci. Sin.-Phys. Mech. Astron.}\ }\textbf {\bibinfo {volume}
  {55}},\ \bibinfo {eid} {2420} (\bibinfo {year} {2012})}\BibitemShut {NoStop}%
\bibitem [{\citenamefont {Zhao}\ \emph {et~al.}(2010)\citenamefont {Zhao},
  \citenamefont {Li}, \citenamefont {Yao},\ and\ \citenamefont
  {Meng}}]{Zhao2010_PRC82-054319}%
  \BibitemOpen
  \bibfield  {author} {\bibinfo {author} {\bibfnamefont {P.~W.}\ \bibnamefont
  {Zhao}}, \bibinfo {author} {\bibfnamefont {Z.~P.}\ \bibnamefont {Li}},
  \bibinfo {author} {\bibfnamefont {J.~M.}\ \bibnamefont {Yao}}, \ and\
  \bibinfo {author} {\bibfnamefont {J.}~\bibnamefont {Meng}},\ }\href {\doibase
  10.1103/PhysRevC.82.054319} {\bibfield  {journal} {\bibinfo  {journal} {Phys.
  Rev. C}\ }\textbf {\bibinfo {volume} {82}},\ \bibinfo {pages} {054319}
  (\bibinfo {year} {2010})}\BibitemShut {NoStop}%
\bibitem [{\citenamefont {Meng}(1993)}]{Meng1993_APS42-368}%
  \BibitemOpen
  \bibfield  {author} {\bibinfo {author} {\bibfnamefont {J.}~\bibnamefont
  {Meng}},\ }\href {http://wulixb.iphy.ac.cn/CN/Y1993/V42/I3/368} {\bibfield
  {journal} {\bibinfo  {journal} {Acta Phys. Sin.}\ }\textbf {\bibinfo {volume}
  {42}},\ \bibinfo {pages} {368} (\bibinfo {year} {1993})}\BibitemShut
  {NoStop}%
\bibitem [{\citenamefont {Frauendorf}\ and\ \citenamefont
  {Meng}(1996)}]{Frauendorf1996_ZPA356-263}%
  \BibitemOpen
  \bibfield  {author} {\bibinfo {author} {\bibfnamefont {S.}~\bibnamefont
  {Frauendorf}}\ and\ \bibinfo {author} {\bibfnamefont {J.}~\bibnamefont
  {Meng}},\ }\href {\doibase 10.1007/BF02769229} {\bibfield  {journal}
  {\bibinfo  {journal} {Z. Phys. A}\ }\textbf {\bibinfo {volume} {356}},\
  \bibinfo {pages} {263} (\bibinfo {year} {1996})}\BibitemShut {NoStop}%
\bibitem [{\citenamefont {Frauendorf}\ and\ \citenamefont
  {Meng}(1997)}]{Frauendorf1997_NPA617-131}%
  \BibitemOpen
  \bibfield  {author} {\bibinfo {author} {\bibfnamefont {S.}~\bibnamefont
  {Frauendorf}}\ and\ \bibinfo {author} {\bibfnamefont {J.}~\bibnamefont
  {Meng}},\ }\href {\doibase 10.1016/S0375-9474(97)00004-3} {\bibfield
  {journal} {\bibinfo  {journal} {Nucl. Phys. A}\ }\textbf {\bibinfo {volume}
  {617}},\ \bibinfo {pages} {131} (\bibinfo {year} {1997})}\BibitemShut
  {NoStop}%
\bibitem [{\citenamefont {Zeng}\ and\ \citenamefont
  {Cheng}(1983)}]{Zeng1983_NPA405-1}%
  \BibitemOpen
  \bibfield  {author} {\bibinfo {author} {\bibfnamefont {J.~Y.}\ \bibnamefont
  {Zeng}}\ and\ \bibinfo {author} {\bibfnamefont {T.~S.}\ \bibnamefont
  {Cheng}},\ }\href {\doibase 10.1016/0375-9474(83)90320-2} {\bibfield
  {journal} {\bibinfo  {journal} {Nucl. Phys. A}\ }\textbf {\bibinfo {volume}
  {405}},\ \bibinfo {pages} {1} (\bibinfo {year} {1983})}\BibitemShut {NoStop}%
\bibitem [{\citenamefont {Zeng}\ \emph
  {et~al.}(1994{\natexlab{a}})\citenamefont {Zeng}, \citenamefont {Jin},\ and\
  \citenamefont {Zhao}}]{Zeng1994_PRC50-1388}%
  \BibitemOpen
  \bibfield  {author} {\bibinfo {author} {\bibfnamefont {J.~Y.}\ \bibnamefont
  {Zeng}}, \bibinfo {author} {\bibfnamefont {T.~H.}\ \bibnamefont {Jin}}, \
  and\ \bibinfo {author} {\bibfnamefont {Z.~J.}\ \bibnamefont {Zhao}},\ }\href
  {\doibase 10.1103/PhysRevC.50.1388} {\bibfield  {journal} {\bibinfo
  {journal} {Phys. Rev. C}\ }\textbf {\bibinfo {volume} {50}},\ \bibinfo
  {pages} {1388} (\bibinfo {year} {1994}{\natexlab{a}})}\BibitemShut {NoStop}%
\bibitem [{\citenamefont {Zeng}\ \emph
  {et~al.}(1994{\natexlab{b}})\citenamefont {Zeng}, \citenamefont {Lei},
  \citenamefont {Jin},\ and\ \citenamefont {Zhao}}]{Zeng1994_PRC50-746}%
  \BibitemOpen
  \bibfield  {author} {\bibinfo {author} {\bibfnamefont {J.~Y.}\ \bibnamefont
  {Zeng}}, \bibinfo {author} {\bibfnamefont {Y.~A.}\ \bibnamefont {Lei}},
  \bibinfo {author} {\bibfnamefont {T.~H.}\ \bibnamefont {Jin}}, \ and\
  \bibinfo {author} {\bibfnamefont {Z.~J.}\ \bibnamefont {Zhao}},\ }\href
  {\doibase 10.1103/PhysRevC.50.746} {\bibfield  {journal} {\bibinfo  {journal}
  {Phys. Rev. C}\ }\textbf {\bibinfo {volume} {50}},\ \bibinfo {pages} {746}
  (\bibinfo {year} {1994}{\natexlab{b}})}\BibitemShut {NoStop}%
\bibitem [{\citenamefont {Liu}\ and\ \citenamefont
  {Zeng}(2002)}]{Liu2002_PRC66-067301}%
  \BibitemOpen
  \bibfield  {author} {\bibinfo {author} {\bibfnamefont {S.~X.}\ \bibnamefont
  {Liu}}\ and\ \bibinfo {author} {\bibfnamefont {J.~Y.}\ \bibnamefont {Zeng}},\
  }\href {\doibase 10.1103/PhysRevC.66.067301} {\bibfield  {journal} {\bibinfo
  {journal} {Phys. Rev. C}\ }\textbf {\bibinfo {volume} {66}},\ \bibinfo
  {pages} {067301} (\bibinfo {year} {2002})}\BibitemShut {NoStop}%
\bibitem [{\citenamefont {Zeng}\ \emph {et~al.}(2001)\citenamefont {Zeng},
  \citenamefont {Liu}, \citenamefont {Lei},\ and\ \citenamefont
  {Yu}}]{Zeng2001_PRC63-024305}%
  \BibitemOpen
  \bibfield  {author} {\bibinfo {author} {\bibfnamefont {J.~Y.}\ \bibnamefont
  {Zeng}}, \bibinfo {author} {\bibfnamefont {S.~X.}\ \bibnamefont {Liu}},
  \bibinfo {author} {\bibfnamefont {Y.~A.}\ \bibnamefont {Lei}}, \ and\
  \bibinfo {author} {\bibfnamefont {L.}~\bibnamefont {Yu}},\ }\href {\doibase
  10.1103/PhysRevC.63.024305} {\bibfield  {journal} {\bibinfo  {journal} {Phys.
  Rev. C}\ }\textbf {\bibinfo {volume} {63}},\ \bibinfo {pages} {024305}
  (\bibinfo {year} {2001})}\BibitemShut {NoStop}%
\bibitem [{\citenamefont {Liu}\ \emph {et~al.}(2002)\citenamefont {Liu},
  \citenamefont {Zeng},\ and\ \citenamefont {Zhao}}]{Liu2002_PRC66-024320}%
  \BibitemOpen
  \bibfield  {author} {\bibinfo {author} {\bibfnamefont {S.~X.}\ \bibnamefont
  {Liu}}, \bibinfo {author} {\bibfnamefont {J.~Y.}\ \bibnamefont {Zeng}}, \
  and\ \bibinfo {author} {\bibfnamefont {E.~G.}\ \bibnamefont {Zhao}},\ }\href
  {\doibase 10.1103/PhysRevC.66.024320} {\bibfield  {journal} {\bibinfo
  {journal} {Phys. Rev. C}\ }\textbf {\bibinfo {volume} {66}},\ \bibinfo
  {pages} {024320} (\bibinfo {year} {2002})}\BibitemShut {NoStop}%
\bibitem [{\citenamefont {Wu}\ \emph {et~al.}(2011)\citenamefont {Wu},
  \citenamefont {Zhang}, \citenamefont {Zeng},\ and\ \citenamefont
  {Lei}}]{Wu2011_PRC83-034323}%
  \BibitemOpen
  \bibfield  {author} {\bibinfo {author} {\bibfnamefont {X.}~\bibnamefont
  {Wu}}, \bibinfo {author} {\bibfnamefont {Z.~H.}\ \bibnamefont {Zhang}},
  \bibinfo {author} {\bibfnamefont {J.~Y.}\ \bibnamefont {Zeng}}, \ and\
  \bibinfo {author} {\bibfnamefont {Y.~A.}\ \bibnamefont {Lei}},\ }\href
  {\doibase 10.1103/PhysRevC.83.034323} {\bibfield  {journal} {\bibinfo
  {journal} {Phys. Rev. C}\ }\textbf {\bibinfo {volume} {83}},\ \bibinfo
  {pages} {034323} (\bibinfo {year} {2011})}\BibitemShut {NoStop}%
\bibitem [{\citenamefont {Liu}\ \emph {et~al.}(2004)\citenamefont {Liu},
  \citenamefont {Zeng},\ and\ \citenamefont {Yu}}]{Liu2004_NPA735-77}%
  \BibitemOpen
  \bibfield  {author} {\bibinfo {author} {\bibfnamefont {S.~X.}\ \bibnamefont
  {Liu}}, \bibinfo {author} {\bibfnamefont {J.~Y.}\ \bibnamefont {Zeng}}, \
  and\ \bibinfo {author} {\bibfnamefont {L.}~\bibnamefont {Yu}},\ }\href
  {\doibase 10.1016/j.nuclphysa.2004.02.007} {\bibfield  {journal} {\bibinfo
  {journal} {Nucl. Phys. A}\ }\textbf {\bibinfo {volume} {735}},\ \bibinfo
  {pages} {77} (\bibinfo {year} {2004})}\BibitemShut {NoStop}%
\bibitem [{\citenamefont {Zhang}\ \emph
  {et~al.}(2009{\natexlab{a}})\citenamefont {Zhang}, \citenamefont {Wu},
  \citenamefont {Lei},\ and\ \citenamefont {Zeng}}]{Zhang2009_NPA816-19}%
  \BibitemOpen
  \bibfield  {author} {\bibinfo {author} {\bibfnamefont {Z.~H.}\ \bibnamefont
  {Zhang}}, \bibinfo {author} {\bibfnamefont {X.}~\bibnamefont {Wu}}, \bibinfo
  {author} {\bibfnamefont {Y.~A.}\ \bibnamefont {Lei}}, \ and\ \bibinfo
  {author} {\bibfnamefont {J.~Y.}\ \bibnamefont {Zeng}},\ }\href {\doibase
  10.1016/j.nuclphysa.2008.10.008} {\bibfield  {journal} {\bibinfo  {journal}
  {Nucl. Phys. A}\ }\textbf {\bibinfo {volume} {816}},\ \bibinfo {pages} {19}
  (\bibinfo {year} {2009}{\natexlab{a}})}\BibitemShut {NoStop}%
\bibitem [{\citenamefont {Zhang}\ \emph
  {et~al.}(2009{\natexlab{b}})\citenamefont {Zhang}, \citenamefont {Lei},\ and\
  \citenamefont {Zeng}}]{Zhang2009_PRC80-034313}%
  \BibitemOpen
  \bibfield  {author} {\bibinfo {author} {\bibfnamefont {Z.~H.}\ \bibnamefont
  {Zhang}}, \bibinfo {author} {\bibfnamefont {Y.~A.}\ \bibnamefont {Lei}}, \
  and\ \bibinfo {author} {\bibfnamefont {J.~Y.}\ \bibnamefont {Zeng}},\ }\href
  {\doibase 10.1103/PhysRevC.80.034313} {\bibfield  {journal} {\bibinfo
  {journal} {Phys. Rev. C}\ }\textbf {\bibinfo {volume} {80}},\ \bibinfo
  {pages} {034313} (\bibinfo {year} {2009}{\natexlab{b}})}\BibitemShut
  {NoStop}%
\bibitem [{\citenamefont {He}\ \emph {et~al.}(2005)\citenamefont {He},
  \citenamefont {Yu}, \citenamefont {Zeng},\ and\ \citenamefont
  {Zhao}}]{He2005_NPA760-263}%
  \BibitemOpen
  \bibfield  {author} {\bibinfo {author} {\bibfnamefont {X.}~\bibnamefont
  {He}}, \bibinfo {author} {\bibfnamefont {S.}~\bibnamefont {Yu}}, \bibinfo
  {author} {\bibfnamefont {J.}~\bibnamefont {Zeng}}, \ and\ \bibinfo {author}
  {\bibfnamefont {E.}~\bibnamefont {Zhao}},\ }\href {\doibase
  10.1016/j.nuclphysa.2005.06.006} {\bibfield  {journal} {\bibinfo  {journal}
  {Nucl. Phys. A}\ }\textbf {\bibinfo {volume} {760}},\ \bibinfo {pages} {263}
  (\bibinfo {year} {2005})}\BibitemShut {NoStop}%
\bibitem [{\citenamefont {Zhang}\ \emph {et~al.}(2011)\citenamefont {Zhang},
  \citenamefont {Zeng}, \citenamefont {Zhao},\ and\ \citenamefont
  {Zhou}}]{Zhang2011_PRC83-011304R}%
  \BibitemOpen
  \bibfield  {author} {\bibinfo {author} {\bibfnamefont {Z.-H.}\ \bibnamefont
  {Zhang}}, \bibinfo {author} {\bibfnamefont {J.-Y.}\ \bibnamefont {Zeng}},
  \bibinfo {author} {\bibfnamefont {E.-G.}\ \bibnamefont {Zhao}}, \ and\
  \bibinfo {author} {\bibfnamefont {S.-G.}\ \bibnamefont {Zhou}},\ }\href
  {\doibase 10.1103/PhysRevC.83.011304} {\bibfield  {journal} {\bibinfo
  {journal} {Phys. Rev. C}\ }\textbf {\bibinfo {volume} {83}},\ \bibinfo
  {pages} {011304R} (\bibinfo {year} {2011})}\BibitemShut {NoStop}%
\bibitem [{\citenamefont {Zhang}\ \emph
  {et~al.}(2012{\natexlab{a}})\citenamefont {Zhang}, \citenamefont {He},
  \citenamefont {Zeng}, \citenamefont {Zhao},\ and\ \citenamefont
  {Zhou}}]{Zhang2012_PRC85-014324}%
  \BibitemOpen
  \bibfield  {author} {\bibinfo {author} {\bibfnamefont {Z.-H.}\ \bibnamefont
  {Zhang}}, \bibinfo {author} {\bibfnamefont {X.-T.}\ \bibnamefont {He}},
  \bibinfo {author} {\bibfnamefont {J.-Y.}\ \bibnamefont {Zeng}}, \bibinfo
  {author} {\bibfnamefont {E.-G.}\ \bibnamefont {Zhao}}, \ and\ \bibinfo
  {author} {\bibfnamefont {S.-G.}\ \bibnamefont {Zhou}},\ }\href {\doibase
  10.1103/PhysRevC.85.014324} {\bibfield  {journal} {\bibinfo  {journal} {Phys.
  Rev. C}\ }\textbf {\bibinfo {volume} {85}},\ \bibinfo {pages} {014324}
  (\bibinfo {year} {2012}{\natexlab{a}})}\BibitemShut {NoStop}%
\bibitem [{\citenamefont {Zhang}\ \emph
  {et~al.}(2012{\natexlab{b}})\citenamefont {Zhang}, \citenamefont {Meng},
  \citenamefont {Zhao},\ and\ \citenamefont
  {Zhou}}]{Zhang2012_arxiv1208.1156v1}%
  \BibitemOpen
  \bibfield  {author} {\bibinfo {author} {\bibfnamefont {Z.-H.}\ \bibnamefont
  {Zhang}}, \bibinfo {author} {\bibfnamefont {J.}~\bibnamefont {Meng}},
  \bibinfo {author} {\bibfnamefont {E.-G.}\ \bibnamefont {Zhao}}, \ and\
  \bibinfo {author} {\bibfnamefont {S.-G.}\ \bibnamefont {Zhou}},\ }\href
  {http://arxiv.org/abs/1208.1156} {\bibfield  {journal} {\bibinfo  {journal}
  {arxiv}\ ,\ \bibinfo {pages} {1208.1156v1}} (\bibinfo {year}
  {2012}{\natexlab{b}})}\BibitemShut {NoStop}%
\bibitem [{\citenamefont {Wu}\ and\ \citenamefont
  {Zeng}(1989)}]{Wu1989_PRC39-666}%
  \BibitemOpen
  \bibfield  {author} {\bibinfo {author} {\bibfnamefont {C.~S.}\ \bibnamefont
  {Wu}}\ and\ \bibinfo {author} {\bibfnamefont {J.~Y.}\ \bibnamefont {Zeng}},\
  }\href {\doibase 10.1103/PhysRevC.39.666} {\bibfield  {journal} {\bibinfo
  {journal} {Phys. Rev. C}\ }\textbf {\bibinfo {volume} {39}},\ \bibinfo
  {pages} {666} (\bibinfo {year} {1989})}\BibitemShut {NoStop}%
\bibitem [{\citenamefont {Meng}\ \emph {et~al.}(2006)\citenamefont {Meng},
  \citenamefont {Guo}, \citenamefont {Liu},\ and\ \citenamefont
  {Zhang}}]{Meng2006_FPC1-38}%
  \BibitemOpen
  \bibfield  {author} {\bibinfo {author} {\bibfnamefont {J.}~\bibnamefont
  {Meng}}, \bibinfo {author} {\bibfnamefont {J.-Y.}\ \bibnamefont {Guo}},
  \bibinfo {author} {\bibfnamefont {L.}~\bibnamefont {Liu}}, \ and\ \bibinfo
  {author} {\bibfnamefont {S.-Q.}\ \bibnamefont {Zhang}},\ }\href {\doibase
  10.1007/s11467-005-0013-5} {\bibfield  {journal} {\bibinfo  {journal}
  {Frontiers Phys. China}\ }\textbf {\bibinfo {volume} {1}},\ \bibinfo {pages}
  {38} (\bibinfo {year} {2006})}\BibitemShut {NoStop}%
\bibitem [{\citenamefont {Pillet}\ \emph {et~al.}(2002)\citenamefont {Pillet},
  \citenamefont {Quentin},\ and\ \citenamefont
  {Libert}}]{Pillet2002_NPA697-141}%
  \BibitemOpen
  \bibfield  {author} {\bibinfo {author} {\bibfnamefont {N.}~\bibnamefont
  {Pillet}}, \bibinfo {author} {\bibfnamefont {P.}~\bibnamefont {Quentin}}, \
  and\ \bibinfo {author} {\bibfnamefont {J.}~\bibnamefont {Libert}},\ }\href
  {\doibase 10.1016/S0375-9474(01)01240-4} {\bibfield  {journal} {\bibinfo
  {journal} {Nucl. Phys. A}\ }\textbf {\bibinfo {volume} {697}},\ \bibinfo
  {pages} {141} (\bibinfo {year} {2002})}\BibitemShut {NoStop}%
\bibitem [{\citenamefont {Molique}\ and\ \citenamefont
  {Dudek}(1997)}]{Molique1997_PRC56-1795}%
  \BibitemOpen
  \bibfield  {author} {\bibinfo {author} {\bibfnamefont {H.}~\bibnamefont
  {Molique}}\ and\ \bibinfo {author} {\bibfnamefont {J.}~\bibnamefont
  {Dudek}},\ }\href {\doibase 10.1103/PhysRevC.56.1795} {\bibfield  {journal}
  {\bibinfo  {journal} {Phys. Rev. C}\ }\textbf {\bibinfo {volume} {56}},\
  \bibinfo {pages} {1795} (\bibinfo {year} {1997})}\BibitemShut {NoStop}%
\bibitem [{\citenamefont {Nilsson}\ \emph {et~al.}(1969)\citenamefont
  {Nilsson}, \citenamefont {Tsang}, \citenamefont {Sobiczewski}, \citenamefont
  {Szymaski}, \citenamefont {Wycech}, \citenamefont {Gustafson}, \citenamefont
  {Lamm}, \citenamefont {M\"{o}ller},\ and\ \citenamefont
  {Nilsson}}]{Nilsson1969_NPA131-1}%
  \BibitemOpen
  \bibfield  {author} {\bibinfo {author} {\bibfnamefont {S.~G.}\ \bibnamefont
  {Nilsson}}, \bibinfo {author} {\bibfnamefont {C.~F.}\ \bibnamefont {Tsang}},
  \bibinfo {author} {\bibfnamefont {A.}~\bibnamefont {Sobiczewski}}, \bibinfo
  {author} {\bibfnamefont {Z.}~\bibnamefont {Szymaski}}, \bibinfo {author}
  {\bibfnamefont {S.}~\bibnamefont {Wycech}}, \bibinfo {author} {\bibfnamefont
  {C.}~\bibnamefont {Gustafson}}, \bibinfo {author} {\bibfnamefont {I.~L.}\
  \bibnamefont {Lamm}}, \bibinfo {author} {\bibfnamefont {P.}~\bibnamefont
  {M\"{o}ller}}, \ and\ \bibinfo {author} {\bibfnamefont {B.}~\bibnamefont
  {Nilsson}},\ }\href {\doibase 10.1016/0375-9474(69)90809-4} {\bibfield
  {journal} {\bibinfo  {journal} {Nucl. Phys. A}\ }\textbf {\bibinfo {volume}
  {131}},\ \bibinfo {pages} {1} (\bibinfo {year} {1969})}\BibitemShut {NoStop}%
\end{thebibliography}


\end{document}